\documentstyle{article}

\newcommand{\zeile}[1]{\vskip #1\baselineskip}

\begin{document}

\pagenumbering{roman}

\section*{Preface}

The history of critical phenomena goes back to the year 1869
when Andrews discovered the critical point of carbon
dioxide, located at about 31\,$^{\circ}$C and $73$ atmospheres pressure.
In the neighborhood of this point the carbon dioxide was observed to
become opalescent, that is, light is strongly scattered. This is
nowadays interpreted as coming from the strong fluctuations of the
system close to the critical point.

Subsequently, a wide variety of physical systems were realized to
display critical points as well. Of particular importance was the
observation of a critical point in ferromagnetic iron by Curie.
Further examples include multicomponent fluids and alloys, superfluids,
superconductors, polymers and may even extend to the quark-gluon
plasma and the early universe as a whole. Early theoretical
investigations tried to reduce the problem to a very small number of
degrees of freedom, such as the van der Waals equation and mean field
approximations and culminating in Landau's general theory of
critical phenomena. In a dramatic development, Onsager's exact
solution of the two-dimensional Ising model made clear the important
role of the critical fluctuations. Their role was taken into account
in the subsequent developments leading to the scaling theories of
critical phenomena and the renormalization group. These developements
have achieved a precise description of the close neighborhood of the
critical point and results are often in good agreement with experiments.
In contrast to the general understanding a century ago,
the presence of fluctuations on all length scales at a critical point
is today emphasized. This can be briefly summarized by saying that
at a critical point a system is {\em scale invariant}.

{\em Conformal invariance} has been known for almost a century in connection
with scale invariance. For example, Maxwell's equations in the vacuum
are scale invariant as well as conformal invariant. This feature arises
generally in quantum field theories with a local energy-momentum tensor.
However, the first application of conformal invariance to critical phenomena
was made only in 1970 by Polyakov. At that time, the consequences of
conformal invariance for an arbitrary number of space dimensions were
considered to be far from spectacular.
This is completely different in two dimensions
as was pointed out in the seminal work by Belavin, Polyakov and
A. Zamolodchikov in 1984. This is because, for two dimensions, the conformal
group is infinite-dimensional and much stronger constraints on the
multipoint correlation functions of the system are obtained.

In a sense, conformal invariance is the logical extension of scale
invariance. Scale invariance requires the invariance of the system
under a {\em uniform} length rescaling. Conformal invariance also
permits a non-uniform, local, rescaling and only requires that angles
are kept unchanged. This extension is in fact very natural since
it can be shown that for
any system which is invariant under translations and rotations, at least
in the continuum limit, is scale invariant and has short-ranged
interactions, conformal invariance follows automatically.

For two-dimensional systems, conformal invariance has considerably
extended our knowledge of the nature of a critical point. While
scale invariance alone is capable of casting systems into universality
classes only dependent on a few selected properties like the global
symmetry, the dimension of the space and of the number of components of
the order parameter, two-dimensional conformal invariance yields a
classification of the critical point partition functions and thereby
furnishes exact values of the critical exponents. Furthermore, the
critical multipoint correlation functions of the local variables
of the system can be determined exactly.

These prospects should place conformal invariance on the center
stage of investigations on critical phenomena. However, the
technical tools required for its understanding are quite elaborate.
One of the central notions, the {\em central charge}, requires some
profound background on anomalies which is not necessarily in the
toolkit of a condensed-matter physicist. Rather, many of the basic
concepts and techniques were developed in the context of string theory
and most of the existing reviews on the subject assume quite some
knowledge on quantum field theory on the side of the reader. On the
other hand, we believe that the time is ready for conformal invariance
techniques to enter condensed matter applications
on a wide front. In writing
this introduction, we have tried to meet the needs of a reader with
some exposure to scaling and the renormalization group without
being an expert in quantum field theory. To do so, we give
a joint presentation of both the
field theory techniques required as well as their explicit application
in lattice systems.
Numerical techniques in connection with finite-lattice
systems will be emphasized, having in mind a reader with a
good physical understanding of the
lattice model who is curious about what conformal invariance
techniques can reveal about its behaviour. At the same time, we have laid
some accent on immediate applications to lattice systems and have
tried to illustrate the phenomenological consequences of conformal
invariance as explicitly as possible. Finite-size effects, rather than
being a nuisance, will appear at a central position throughout.

This book has grown out of a joint two-trimester course held at the
University of Fribourg in the winter 1991/1992. We hope it
may serve as a first glance into the field and will prepare the
reader for the study of more advanced presentations, some of which
are given in the general references. In selecting the material
to be presented in a first introduction to conformal invariance, we had
to restrict ourselves to the basic foundations of the theory and many of the
more advanced applications had regretably
to be left out completely. Although string
theory stands at the origin of conformal invariance, no mention is
made of it here. While conformal invariance provides one of the
building blocks of strings in higher dimensions, it has also led
to profound studies of field theories on a fluctuating metric. These
studies include pure two-dimensional quantum gravity as well as
spin systems on lattices with random connectivity, rewritten in the form of
matrix models. We skip completely the
fascinating subject of exactly integrable systems with the exceptions only
of the two-dimensional Ising model and a brief sketch of A. Zamolodchikov's
theory of two-dimensional systems perturbed away from their critical point.
The most spectacular result of this has been the proof that the two-dimensional
Ising model at its critical temperature, but in a magnetic field, is
integrable.
These developments have also stimulated further investigations in
mathematics, uncovering for example very deep and interesting relationships
to the theory of knots and links. We merely mention here that
polynomial invariants of knots have reappeared in connection with the
partition functions of two-dimensional integrable systems. Conformal
invariance has also added to the general understanding of the flow
under the renormalisation group, referred to as the $c$-theorem. We
restrict ourselves
here exclusively to its occurrence in two dimensions and do not
go into the existing generalisations in four dimensions.
Since conformal invariance acts primarily as a dynamical symmetry which
allows one to write the spectrum of the transfer matrix in terms of the
irreducible representations of the conformal algebra, it is worth
looking for conformal symmetries more general than a Lie group.
Of the extensions of the conformal group we only briefly mention $N=1$
superconformal invariance and skip higher superconformal algebras as well
as $W$ algebras, braid groups and quantum groups. The only Kac-Moody
algebra treated here in any detail is the $U(1)$, including its
shifted representations. Temperley-Lieb algebras are merely mentioned
although they provide but the first example of a whole class of
new symmetry structures present in conformally
invariant and integrable systems.
On the side of the more advanced applications, we do not cover
weakly disordered systems, polymers and random walks, the quantum Hall
effect or the Kondo effect. It would be tempting to use conformal invariance
for a better understanding of high $T_c$ superconductivity. Conformal
invariance techniques have also been useful in calculating exactly
the fractal dimensions of clusters as defined for example by the order
parameter of spin systems. For an introduction to these active
reasearch topics, see the general references in the bibliography or the
references to review articles in the text.

In chapter~1, we recall some well-known
notions of scaling relevant in connection with their subsequent generalisation
to conformal invariance. We also repeat the correspondence between
quantum field theory and classical statistical equilibrium mechanics.
This is formulated via the {\em transfer matrix}, which will play a major role
in what follows. In chapters~2 to 7, we then give
the field theory point of view of conformal invariance. We shall show
how this can be used to calculate explicitly the critical multipoint
correlation functions and shall go through the example of the Ising model
in full detail. Following our two-pronged approach, we present at an
early stage (chapter~3) the theory of finite-size scaling and
the important contributions to it from conformal invariance. Conversely, the
new conformal results provide a simple and efficient means for the calculation
of critical exponents and the central charge from a given lattice system.
We then turn to lattice systems. A large part of the original work done on
conformal invariance uses
as a simplifying technical device an extremely anisotropic limit of the
transfer matrix, which has the virtue of turning a fully populated matrix into
a sparse one. This is quite useful for numerical calculations and also
sheds more light on universality, as detailed in chapter~8.
We describe the numerical techniques needed with an accent
laid on the subleties of
finite-lattice extrapolation in chapter~9.
The Ising model example is then studied
again in chapter~10 to show how a
lattice system can be treated from the start from the point of
view of conformal invariance.

The subsequent chapters deal with more advanced applications. Modular
invariance is treated in chapter~11 and is shown to lead to
a classification of the critical point partition functions. Examples
beyond the Ising model, along with additional concepts, are presented
in chapter~12. In chapters~13 and 14, we consider
the effects of both relevant and irrelevant perturbing operators, culminating
in the beautiful developments involing $S$-matrix theory which led
to the recognition of
the integrability of the two-dimensional Ising model
in a magnetic field. Surface critical
phenomena are treated in chapter~15 and we close with an outlook
towards possible applications in critical dynamics.

To keep track of every new piece of work in this rapidly expanding
field has been beyond our capabilites. We compiled in the bibliography
the references we used in writing this text and added some more intended
as suggestions for further reading. The first half of the general references
gives introductory texts on critical phenomena, followed by
earlier reviews on conformal invariances and a few suggestions for catching
up with the current lines of research. The selection
is certainly incomplete and we sincerely
apologize to any author who might feel that we did not give proper credit to
his contribution to the field. Inevitably the bibiography
reflects our interests and/or our lack of knowledge of some directions
under the conformal umbrella. Citations in the text are often in historical
order.

Finally, we have the pleasant task of thanking all those who have contributed
to making this work possible.
We thank Prof. D. Baeriswyl for his kind invitation to give the lectures
from which this book has grown.
We are deeply indebted to R. Flume and V. Rittenberg for introducing us
to the subject and for continuous support and encouragement. One
or other or both of us have received advise or support from and/or
had the pleasure to collaborate with
J.L. Cardy, E. Domany, M. Droz, L. Frachebourg, G.v. Gehlen,
H.J. Herrmann, A.W.W. Ludwig, M.J. Martins, G. Mussardo, A. Patk\'os,
R. Peschanski, V. Privman, F. Ravanini, H. Saleur, G. Sch\"utz,
N.M. \v Svraki\'c, R.A. Weston and J.-B. Zuber.
Our warmest thanks go to all of them.

We thank the Institut f\"ur Theoretische Physik of the Universit\"at Bern and
the D\'epartement de Physique Th\'eorique of the Universit\'e de Gen\`eve for
support. This work was supported in part by the Swiss National Science
Foundation.

\zeile{2}
\noindent{\it Bern and Gen\`eve} \hfill Philippe Christe \\
{\it January 1993} \hfill Malte Henkel

\newpage


\contentsline {section}{List of tables}{xiii}
\contentsline {section}{List of figures}{xv}
\contentsline {section}{\numberline {1}Critical Phenomena: a Reminder}{1}
\contentsline {subsection}{\numberline {1.1}Phase diagrams and critical
exponents}{1}
\contentsline {subsection}{\numberline {1.2}Scaling relations}{4}
\contentsline {subsection}{\numberline {1.3}Some simple spin systems}{7}
\contentsline {subsubsection}{The Yang-Lee edge singularity.}{8}
\contentsline {subsubsection}{The tricritical Ising model.}{9}
\contentsline {subsubsection}{The $p$-state Potts model.}{10}
\contentsline {subsubsection}{The vector Potts model.}{14}
\contentsline {subsubsection}{The XY model.}{15}
\contentsline {subsubsection}{Restricted solid-on-solid models.}{16}
\contentsline {subsection}{\numberline {1.4}Correspondence between
statistical systems and field theory}{18}
\contentsline {subsection}{\numberline {1.5}Correspondence of physical
quantities}{21}
\contentsline {subsubsection}{Free energy density.}{21}
\contentsline {subsubsection}{Correlation functions.}{22}
\contentsline {subsubsection}{Correlation lengths.}{22}
\contentsline {section}{\numberline {2}Conformal Invariance and the
Stress-Energy Tensor}{24}
\contentsline {subsection}{\numberline {2.1}Conformal group}{24}
\contentsline {subsection}{\numberline {2.2}Conformal algebra in two
dimensions}{25}
\contentsline {subsection}{\numberline {2.3}Conformal theory in two
dimensions}{27}
\contentsline {subsection}{\numberline {2.4}Elementary correlation
functions}{28}
\contentsline {subsection}{\numberline {2.5}The stress-energy tensor}{30}
\contentsline {section}{\numberline {3}Finite Size Scaling}{38}
\contentsline {subsection}{\numberline {3.1}Statistical systems in finite
geometries}{38}
\contentsline {subsection}{\numberline {3.2}Finite-size scaling hypothesis}{39}
\contentsline {subsection}{\numberline {3.3}Universality}{42}
\contentsline {subsection}{\numberline {3.4}Phenomenological
renormalization}{45}
\contentsline {subsection}{\numberline {3.5}Consequences of conformal
invariance}{47}
\contentsline {section}{\numberline {4}Representation Theory of the Virasoro
Algebra}{51}
\contentsline {subsection}{\numberline {4.1}Verma module}{51}
\contentsline {subsection}{\numberline {4.2}Hilbert space structure}{54}
\contentsline {subsection}{\numberline {4.3}Null-vectors}{56}
\contentsline {subsection}{\numberline {4.4}Null-vectors and correlation
functions}{57}
\contentsline {subsection}{\numberline {4.5}Kac formula and unitarity}{60}
\contentsline {subsection}{\numberline {4.6}Minimal characters}{62}
\contentsline {section}{\numberline {5}Operator Algebra and Correlation
Functions}{65}
\contentsline {subsection}{\numberline {5.1}Operator algebra and
associativity}{65}
\contentsline {subsection}{\numberline {5.2}Analyticity and the monodromy
problem}{70}
\contentsline {subsection}{\numberline {5.3}The Riemann point of view}{72}
\contentsline {section}{\numberline {6}The Ising Model Correlation
Functions}{76}
\contentsline {subsection}{\numberline {6.1}Four-point function of the
spin-density operator}{76}
\contentsline {subsection}{\numberline {6.2}Four-point function of the
energy-density operator}{79}
\contentsline {subsection}{\numberline {6.3}Mixed four-point function}{81}
\contentsline {subsection}{\numberline {6.4}Semi-local four-point
functions}{82}
\contentsline {section}{\numberline {7}Coulomb Gas Realization}{84}
\contentsline {subsection}{\numberline {7.1}Free massless boson gas}{84}
\contentsline {subsection}{\numberline {7.2}Screened Coulomb gas}{87}
\contentsline {subsection}{\numberline {7.3}Minimal correlation functions}{89}
\contentsline {subsection}{\numberline {7.4}Minimal algebras and OPE
coefficients}{91}
\contentsline {section}{\numberline {8}The Hamiltonian Limit and
Universality}{94}
\contentsline {subsection}{\numberline {8.1}Hamiltonian limit in the Ising
model}{94}
\contentsline {subsection}{\numberline {8.2}Hubbard-Stratonovi{\accent 19 c}
transformation}{97}
\contentsline {subsection}{\numberline {8.3}Hamiltonian limit of the
scalar $\phi ^4$ theory}{98}
\contentsline {subsection}{\numberline {8.4}Hamiltonian spectrum and
conformal invariance}{101}
\contentsline {subsection}{\numberline {8.5}Temperley-Lieb algebra}{101}
\contentsline {subsection}{\numberline {8.6}Laudau-Ginzburg
classification}{105}
\contentsline {section}{\numberline {9}Numerical Techniques}{107}
\contentsline {subsection}{\numberline {9.1}Simple properties of quantum
Hamiltonians}{107}
\contentsline {subsection}{\numberline {9.2}Some further physical quantities
and their critical exponents}{109}
\contentsline {subsection}{\numberline {9.3}Translation invariance}{111}
\contentsline {subsection}{\numberline {9.4}Diagonalization}{112}
\contentsline {subsection}{\numberline {9.5}Extrapolation}{115}
\contentsline {subsubsection}{VBS algorithm.}{118}
\contentsline {subsubsection}{BST algorithm.}{119}
\contentsline {section}{\numberline {10}Conformal Invariance in the Ising
Quantum Chain}{122}
\contentsline {subsection}{\numberline {10.1}Exact diagonalization}{122}
\contentsline {subsubsection}{General remarks.}{122}
\contentsline {subsubsection}{Jordan-Wigner transformation.}{123}
\contentsline {subsubsection}{Diagonalization of a quadratic form.}{124}
\contentsline {subsubsection}{Eigenvalue spectrum and normalization.}{125}
\contentsline {subsection}{\numberline {10.2}Character functions}{126}
\contentsline {subsection}{\numberline {10.3}Finite-size scaling analysis}{128}
\contentsline {subsubsection}{Ground state energy.}{128}
\contentsline {subsubsection}{Operator content.}{130}
\contentsline {subsubsection}{Finite-size corrections.}{132}
\contentsline {subsubsection}{Finite-size scaling functions.}{133}
\contentsline {subsection}{\numberline {10.4}The Virasoro generators}{134}
\contentsline {subsection}{\numberline {10.5}Recapitulation}{135}
\contentsline {section}{\numberline {11}Modular Invariance}{137}
\contentsline {subsection}{\numberline {11.1}The modular group}{137}
\contentsline {subsection}{\numberline {11.2}Implementation for minimal
models}{137}
\contentsline {subsection}{\numberline {11.3}Modular invariance at $c=1$}{143}
\contentsline {subsubsection}{ Circle or Coulomb models.}{143}
\contentsline {subsubsection}{ Orbifold models.}{145}
\contentsline {section}{\numberline {12}Further Developments and
Applications}{148}
\contentsline {subsection}{\numberline {12.1}Three-states Potts model}{148}
\contentsline {subsection}{\numberline {12.2}Tricritical Ising model}{149}
\contentsline {subsubsection}{Operator content.}{149}
\contentsline {subsubsection}{Supersymmetry and superconformal
invariance.}{152}
\contentsline {subsection}{\numberline {12.3}Yang-Lee edge singularity}{154}
\contentsline {subsection}{\numberline {12.4}Ashkin-Teller model}{157}
\contentsline {subsubsection}{Relation with the XXZ quantum chain.}{158}
\contentsline {subsubsection}{Global symmetry and boundary conditions.}{158}
\contentsline {subsubsection}{Phase diagram.}{159}
\contentsline {subsubsection}{Operator content on the $c=1$ line.}{160}
\contentsline {subsection}{\numberline {12.5}XY model}{162}
\contentsline {subsection}{\numberline {12.6}XXZ quantum chain}{164}
\contentsline {subsection}{\numberline {12.7}A few remarks on 3$D$
systems}{168}
\contentsline {section}{\numberline {13}Conformal Perturbation Theory}{171}
\contentsline {subsection}{\numberline {13.1}Correlation functions in the
strip geometry}{171}
\contentsline {subsection}{\numberline {13.2}General remarks on corrections
to the critical behaviour}{172}
\contentsline {subsection}{\numberline {13.3}Finite-size corrections}{174}
\contentsline {subsubsection}{Application to the Ising model.}{176}
\contentsline {subsubsection}{Application to the three-states Potts
model.}{177}
\contentsline {subsubsection}{Checking the operator content from finite-size
corrections.}{179}
\contentsline {subsection}{\numberline {13.4}Finite-size scaling
functions}{179}
\contentsline {subsubsection}{Ising model: thermal perturbation.}{180}
\contentsline {subsubsection}{Ising model: magnetic perturbation.}{181}
\contentsline {subsection}{\numberline {13.5}Truncation method}{183}
\contentsline {section}{\numberline {14}The Vicinity of the Critical
Point}{187}
\contentsline {subsection}{\numberline {14.1}The c-theorem}{187}
\contentsline {subsection}{\numberline {14.2}Conserved currents close to
criticality}{190}
\contentsline {subsection}{\numberline {14.3}Exact $S$-matrix approach}{193}
\contentsline {subsection}{\numberline {14.4}Phenomenological
consequences}{200}
\contentsline {subsubsection}{Integrable perturbations.}{200}
\contentsline {subsubsection}{Non-integrable perturbations.}{203}
\contentsline {subsection}{\numberline {14.5}Thermodynamic Bethe Ansatz}{207}
\contentsline {subsection}{\numberline {14.6}Asymptotic finite-size scaling
functions}{213}
\contentsline {section}{\numberline {15}Surface Critical Phenomena}{217}
\contentsline {subsection}{\numberline {15.1}Systems with a boundary}{217}
\contentsline {subsection}{\numberline {15.2}Conformal invariance close to a
free surface}{221}
\contentsline {subsection}{\numberline {15.3}Finite-size scaling with free
boundary conditions}{224}
\contentsline {subsection}{\numberline {15.4}Surface operator content}{225}
\contentsline {subsubsection}{Ising model.}{225}
\contentsline {subsubsection}{Three-states Potts model.}{228}
\contentsline {subsubsection}{Tricritical Ising model.}{230}
\contentsline {subsubsection}{Yang-Lee edge singularity.}{230}
\contentsline {subsubsection}{Ashkin-Teller model.}{231}
\contentsline {subsubsection}{XXZ quantum chain.}{232}
\contentsline {subsection}{\numberline {15.5}Defect lines}{233}
\contentsline {section}{\numberline {16}Outlook: Beyond the Conformal
Group}{242}
\contentsline {section}{References}{247}
\contentsline {section}{Index}{256}

\newpage

\section*{List of Tables}

\contentsline {table}{\numberline
{1}{\ignorespaces Possible melting transitions of adatoms. }}{14}
\contentsline {table}{\numberline
{2}{\ignorespaces Analogies between statistical mechanics and quantum
theory. }}{20}
\contentsline {table}{\numberline
{3}{\ignorespaces Correspondence of physical quantities of statistical and
quantum systems. }}{21}
\contentsline {table}{\numberline
{4}{\ignorespaces Bulk scaling and finite-size scaling close to $T_c$
with $\mathaccent "707E {z}=0$. }}{42}
\contentsline {table}{\numberline
{5}{\ignorespaces Kac table for the Ising model and for the $A_5$
RSOS model.}}{61}
\contentsline {table}{\numberline
{6}{\ignorespaces Correspondence between conformal primary fields and the LGW
fields $\phi ^{n}$. }}{105}
\contentsline {table}{\numberline
{7}{\ignorespaces Symmetry properties of a $p$-state $Z_p$ symmetric quantum
Hamiltonian. }}{109}
\contentsline {table}{\numberline
{8}{\ignorespaces Convergence of the Lanczos algorithm.}}{114}
\contentsline {table}{\numberline
{9}{\ignorespaces Finite-lattice extrapolation with the VBS algorithm,
$\alpha =-1$. }}{121}
\contentsline {table}{\numberline
{10}{\ignorespaces Finite-lattice extrapolation with the BST algorithm,
$\omega =1.94$. }}{121}
\contentsline {table}{\numberline
{11}{\ignorespaces Calculation of the degeneracies $d(0,I)$ of the operator
$\phi _{1,1}$ for $m=3$. }}{127}
\contentsline {table}{\numberline
{12}{\ignorespaces Virasoro characters of the primary operators $\phi _{r,s}$
for $m=3$. }}{128}
\contentsline {table}{\numberline
{13}{\ignorespaces Critical point partition functions as given by the
$A$, $D$ and $E$ series of unitary models. }}{140}
\contentsline {table}{\numberline
{14}{\ignorespaces Critical point partition function for $Z_2$ symmetric
systems with antiperiodic boundary conditions in the $A$ series. }}{142}
\contentsline {table}{\numberline
{15}{\ignorespaces Virasoro characters of the primary fields $\phi _{r,s}$
for $m=5$. }}{149}
\contentsline {table}{\numberline
{16}{\ignorespaces Virasoro characters of the primary fields $\phi _{r,s}$
for $m=4$. }}{151}
\contentsline {table}{\numberline
{17}{\ignorespaces Location of the Yang-Lee singularity and conformal
normalization. }}{154}
\contentsline {table}{\numberline
{18}{\ignorespaces Virasoro characters of the primary fields $\phi _{r,s}$
for $p=5,p'=2$. }}{155}
\contentsline {table}{\numberline
{19}{\ignorespaces Sector equivalence of the XXZ chain with the Ashkin-Teller
chain. }}{158}
\contentsline {table}{\numberline
{20}{\ignorespaces Symmetries of the Ashkin-Teller model in dependence of the
boundary conditions. }}{159}
\contentsline {table}{\numberline
{21}{\ignorespaces Operator content of the Ashkin-Teller model for the
$D_4$-invariant boundary conditions $\Sigma ^{0}$, $\Sigma ^{2}$. }}{162}
\contentsline {table}{\numberline
{22}{\ignorespaces Scalar primary fields for periodic boundary conditions of
the Ashkin-Teller quantum chain. }}{162}
\contentsline {table}{\numberline
{23}{\ignorespaces Exponent $\eta (\lambda )$ for the $(1+1)D$ XY model.
}}{164}
\contentsline {table}{\numberline
{24}{\ignorespaces Critical point $t_c$ and amplitude ratio $\Xi $ in the
$(2+1)D$ Ising model. }}{169}
\contentsline {table}{\numberline
{25}{\ignorespaces Finite-size correction coefficients for the three-states
Potts model.}}{177}
\contentsline {table}{\numberline
{26}{\ignorespaces Quasiprimary states up to level four for the Yang-Lee
singularity. }}{184}
\contentsline {table}{\numberline
{27}{\ignorespaces The normalized scalar states and their scaling dimensions
for the truncated space up to level 3. }}{184}
\contentsline {table}{\numberline
{28}{\ignorespaces Counting criterion for the Ising model perturbed with
$\phi _{1,2}$.}}{192}
\contentsline {table}{\numberline
{29}{\ignorespaces Mass ratios $r_i = m_{i+1} / m_1$ for the Ashkin-Teller
model.}}{206}
\contentsline {table}{\numberline
{30}{\ignorespaces Numerical values for $\mu _{abc}$ and
$\rho _{abc} := \lambda _{abc}^{2} / (8 m_{a}^{2} \pcomma \mu _{abc} )$
for the $2D$ Ising model. }}{214}
\contentsline {table}{\numberline
{31}{\ignorespaces Mass shifts $\delta G_i$ for the Ising model in a magnetic
field. }}{216}

\newpage

\section*{List of Figures}

\contentsline {figure}{\numberline {1}{\ignorespaces Typical phase diagram of
a ferromagnet.}}{1}
\contentsline {figure}{\numberline {2}{\ignorespaces Magnetization as
function of $B$ and $T$. }}{2}
\contentsline {figure}{\numberline {3}{\ignorespaces Droplet picture
(schematic).}}{2}
\contentsline {figure}{\numberline {4}{\ignorespaces Zeroes of the Ising
model partition function at $T> T_c$.}}{8}
\contentsline {figure}{\numberline {5}{\ignorespaces Phase diagram of the
tricritical Ising model.}}{10}
\contentsline {figure}{\numberline {6}{\ignorespaces Example of a closed
graph in the $p$-state Potts model. }}{11}
\contentsline {figure}{\numberline {7}{\ignorespaces Plaquette of a square
lattice.}}{16}
\contentsline {figure}{\numberline {8}{\ignorespaces The six possible
plaquettes.}}{17}
\contentsline {figure}{\numberline {9}{\ignorespaces Classical path. }}{19}
\contentsline {figure}{\numberline {10}{\ignorespaces Hypercubic lattice
with "time" and "space" directions. }}{20}
\contentsline {figure}{\numberline {11}{\ignorespaces Conformal transformation
in $d$ dimensions. }}{24}
\contentsline {figure}{\numberline {12}{\ignorespaces Bounded coordinate
transformation. }}{31}
\contentsline {figure}{\numberline {13}{\ignorespaces Convention for the
contours in the definition of the Virasoro generators. }}{35}
\contentsline {figure}{\numberline {14}{\ignorespaces Commutation contours
for the Virasoro algebra. }}{35}
\contentsline {figure}{\numberline {15}{\ignorespaces Specific heat as a
function of temperature for two lattices of sizes $L_1 > L_2$. }}{39}
\contentsline {figure}{\numberline {16}{\ignorespaces Solving the
phenomenological renormalization condition to find $T^*$. }}{46}
\contentsline {figure}{\numberline {17}{\ignorespaces Logarithmic
transformation. }}{48}
\contentsline {figure}{\numberline {18}{\ignorespaces Deformation of a
oriented contour: $\ointop \nolimits _{C_z} = \ointop
\nolimits _C - \sum _i\pcomma \ointop \nolimits _{C_i}$. }}{58}
\contentsline {figure}{\numberline {19}{\ignorespaces Sequence of Verma
submodules.}}{62}
\contentsline {figure}{\numberline {20}{\ignorespaces Crossing symmetry
condition for the four-point functions. }}{69}
\contentsline {figure}{\numberline {21}{\ignorespaces One possible
decomposition of a $n$-point function. }}{69}
\contentsline {figure}{\numberline {22}{\ignorespaces Monodromy
transformations.}}{70}
\contentsline {figure}{\numberline {23}{\ignorespaces Effective potential
$V(\phi )$. }}{99}
\contentsline {figure}{\numberline {24}{\ignorespaces Relationship between
systems in the Ising universality class. }}{100}
\contentsline {figure}{\numberline {25}{\ignorespaces Finite-size estimates
of the exponent $x_{\varepsilon }$ in the $(1+1)D$ Ising model. }}{116}
\contentsline {figure}{\numberline {26}{\ignorespaces Finite-size estimates
for $\beta _c$ in the $3D$ spherical model. }}{117}
\contentsline {figure}{\numberline {27}{\ignorespaces Critical finite-size
scaling spectrum for $H_{0}^{(0)}$ in the $2D$ Ising model. }}{131}
\contentsline {figure}{\numberline {28}{\ignorespaces The torus and the
modular transformations $T$ and $S$. }}{138}
\contentsline {figure}{\numberline {29}{\ignorespaces The manifold $S^1$ and
the orbifold $S^{1}/Z_2$. }}{145}
\contentsline {figure}{\numberline {30}{\ignorespaces Known conformal modular
invariant systems with $c=1$. }}{146}
\contentsline {figure}{\numberline {31}{\ignorespaces Phase diagram of the
$2D$ Ashkin-Teller quantum chain. }}{160}
\contentsline {figure}{\numberline {32}{\ignorespaces Finite-size scaling
functions for the quantum Ising chain.}}{186}
\contentsline {figure}{\numberline {33}{\ignorespaces Mapping the $s$-plane
to the $\theta $-plane.}}{194}
\contentsline {figure}{\numberline {34}{\ignorespaces Three-particle elastic
scattering factorization.}}{195}
\contentsline {figure}{\numberline {35}{\ignorespaces Anomalous
threshold.}}{196}
\contentsline {figure}{\numberline {36}{\ignorespaces Bootstrap
equation.}}{196}
\contentsline {figure}{\numberline {37}{\ignorespaces Mass ratios
$r_i=(\xi _{i+1}/\xi _{1})^{-1}$ for the Ising model in a magnetic field as
a function of $\mu $. }}{202}
\contentsline {figure}{\numberline {38}{\ignorespaces Mass ratios
$r_i=(\xi _{i+1}/\xi _{1})^{-1}$ as functions of $\mu $ for the
tricritical Ising model. }}{204}
\contentsline {figure}{\numberline {39}{\ignorespaces Relation between $Z_2$
breaking perturbations $\phi _{2,2}$ in multicritical Ising models and
the Ashkin-Teller model.}}{206}
\contentsline {figure}{\numberline {40}{\ignorespaces Mass shift in the
$(1+1)D$ Ising model with a magnetic field.}}{215}
\contentsline {figure}{\numberline {41}{\ignorespaces Schematic local order
parameter profiles.}}{218}
\contentsline {figure}{\numberline {42}{\ignorespaces Schematic surface phase
diagram. }}{219}
\contentsline {figure}{\numberline {43}{\ignorespaces Scaled spectrum
$x(\kappa )$ of the $Q=0$ sector of the Ising quantum Hamiltonian with a
single defect line.}}{236}
\contentsline {figure}{\numberline {44}{\ignorespaces Scaled spectrum
$x(\kappa )$ of the $Q=1$ sector of the Ising quantum Hamiltonian with
a single defect line.}}{237}

\end{document}